\documentclass[%
 reprint,
 amsmath,amssymb,
 aps,
]{revtex4-2}

\usepackage{graphicx}
\usepackage{dcolumn}
\usepackage{bm}
\usepackage{amsthm}
\newtheorem*{proposition}{Proposition}

\begin{document}

\title{Local Geometric Bounds on\\Generalized Entropy Evolution along Null Horizons}

\author{Erik Bertram}
 \email{mail@erik-bertram.de}
\affiliation{
 Hochschule Fresenius Heidelberg\\
 Sickingenstraße 63-65, 69126 Heidelberg, Germany
}

\date{\today}

\begin{abstract}
We develop a local and covariant framework for constraining the evolution of generalized entropy along null horizons, combining classical geometric methods with constraints from quantum field theory. Building on the quantum focusing conjecture (QFC), which restricts entropy evolution along null directions, we derive a Raychaudhuri-type differential inequality for the generalized expansion that makes its local dependence on expansion and shear explicit. The resulting relation, $\frac{d\Theta}{d\lambda} \le -\theta \Theta + \tfrac{1}{2}\theta^2 - \sigma^2$, reveals a direct interplay between expansion and shear in controlling entropy flow: shear contributes negatively and yields a monotonic suppression of the generalized expansion in shear-dominated regimes, while expansion provides a competing geometric source term. This structure provides a local geometric formulation consistent with the QFC and clarifies its interpretation as a constraint on entropy evolution. We further show that quantum extremal surfaces correspond to configurations characterized by $\Theta=0$, whose local stability properties are governed by the same geometric data. In addition, we derive a bound on the exponential separation of nearby null generators, indicating that the same combination of expansion and shear also controls geometric instability. Our results provide a local geometric refinement of entropy bounds in semiclassical gravity and establish a direct connection between null geometry, quantum energy conditions, and entropy flow, offering a local geometric perspective on horizon thermodynamics beyond global formulations.
\end{abstract}


\maketitle

\section{Introduction}

The interplay between gravitation, thermodynamics, and quantum information has been a central theme in theoretical physics since the discovery that black holes possess entropy and temperature. The identification of horizon entropy with the Bekenstein--Hawking area law \citep{Bekenstein1973,Hawking1975} established a deep connection between geometry and thermodynamics, suggesting that spacetime itself may admit an underlying statistical or information-theoretic description.

Subsequent developments have extended this perspective to more general settings. In particular, the concept of generalized entropy, which combines the geometric area term with the von Neumann entropy of quantum fields, has played a crucial role in understanding semiclassical gravity \citep{Sorkin1983,Bombelli1986,Wall2012}. This framework has led to important results such as the generalized second law and, more recently, quantum energy conditions that constrain the behavior of quantum matter in curved spacetime.

A key advance in this direction is the quantum null energy condition (QNEC), which provides a local inequality relating the second variation of entanglement entropy to the expectation value of the stress-energy tensor \citep{Bousso2016QNEC,Koeller2016,Balakrishnan2019}. The QNEC has been shown to hold in a wide class of quantum field theories and has deep connections to holography, quantum information, and the structure of spacetime itself, making it a natural tool for studying the local interplay between geometry, energy conditions, and quantum entanglement in gravitational systems.

In recent years, these ideas have been further developed in the context of semiclassical gravity and quantum information, leading to a more refined understanding of the interplay between geometry and entropy. In particular, the quantum focusing conjecture (QFC) extends the classical focusing theorem by incorporating the generalized entropy into the evolution of null congruences, suggesting that entropy itself obeys geometric constraints analogous to those of area \citep{Bousso2016}. Closely related developments include the formulation of quantum extremal surfaces (QES) and the emergence of island prescriptions (see, e.g., \citep{Hashimoto2020,Ling2021,Wang2021}), which have provided new insights into the fine-grained entropy of black holes and the structure of spacetime \citep{Engelhardt2015,Almheiri2019,Penington2020}.

At the same time, the classical dynamics of null congruences, governed by the Raychaudhuri equation, provides a geometric description of how bundles of null generators evolve under the influence of expansion, shear, and curvature \citep{Raychaudhuri1955,Wald1984,Poisson2004,Sayan2007}. This equation underlies many fundamental results in general relativity, including the singularity theorems \cite{Penrose1965,Hawking1970,Hawking1972}, and encodes the focusing properties of null geodesics and how they separate from each other. This evolution motivates the introduction of an effective geometric instability exponent, characterizing the sensitivity to initial conditions along the horizon. Previous works have explored the role of such Lyapunov bounds in the context of holographic chaos \citep{Maldacena2016, Blake2016}, and their relation to entropy and scrambling times \citep{Shenker2014}.

Despite these developments, much of the existing work has focused on global or integrated statements, while a fully local differential understanding of entropy evolution along null horizons remains less explored. In particular, it is natural to ask how the geometric data of a null congruence directly control the local flow of generalized entropy. While powerful constraints such as the QFC provide inequalities governing the evolution of entropy, a local differential equation analogous to the Raychaudhuri equation for geometric expansion has so far been lacking. In this sense, existing results constrain but do not dynamically determine the entropy flow.

In this work, we investigate the dynamical evolution of generalized entropy on null horizons by combining the geometric framework of null congruences with constraints from quantum field theory. Our approach thereby focuses on a local notion of entropy flow defined along the horizon generators, encoded in the generalized expansion $\Theta = \frac{1}{dA}\,\mathcal{L}_k dS_{\mathrm{gen}}$, which provides a pointwise characterization of entropy dynamics. However, while the QFC constrains entropy evolution through a monotonicity condition, we derive a more explicit local differential inequality: a Raychaudhuri-type differential inequality that directly relates entropy flow to geometric data of null congruences. This structure naturally admits an interpretation in terms of a dissipative-like entropy flow: rather than leading to classical focusing, it suggests an effective relaxation behavior in shear-dominated regimes. However, we note that throughout this work, interpretations in terms of dissipation or relaxation always refer to the behavior of solutions that saturate or closely approach the derived inequality limit, and should not be understood as exact dynamical evolution laws.

Furthermore, we identify a distinguished combination of expansion and shear, $\mathcal{K} = \tfrac{1}{2}\theta^2 - \sigma^2$, which governs the local behavior of entropy flow and determines whether the system is expansion- or shear-dominated. In shear-dominated regimes, the inequality reduces to a local form consistent with the Quantum Focussing Conjecture.

We further show that quantum extremal surfaces correspond to configurations characterized by $\Theta = 0$, whose existence is controlled by the same geometric data. A linear perturbation analysis indicates damping of small perturbations in the generalized expansion, with shear acting as a dissipative mechanism that suppresses anisotropic fluctuations.

Finally, we explore the connection between shear and the separation of nearby null generators. Using the geodesic deviation equation, we derive a bound on an effective geometric instability exponent, suggesting that the same geometric structures underlie both entropy evolution and separation on null horizons. In stationary black hole and cosmological horizons, this yields a scaling behavior qualitatively consistent with known relations between chaos and temperature.

The remainder of this paper is organized as follows. In Sec.~II, we introduce the geometric setup and define the local entropy density and its evolution. In Sec.~III, we analyze the behavior of nearby null generators using the geodesic deviation equation and derive a bound on an effective geometric instability exponent. We then relate this geometric instability to the obtained entropy inequality. In Sec.~IV, we present applications to black hole and cosmological horizons. We conclude in Sec.~V.

\section{Geometric Setup and Entropy Evolution}

In this section, we introduce the geometric framework underlying our analysis and define a local notion of entropy flow along null generators. Combining the Raychaudhuri equation with the QNEC, we derive a differential inequality governing the evolution of the generalized expansion $\Theta$. This yields the central relation linking null geometry and entropy dynamics.

\subsection{Null Congruences and Kinematics}

We start by considering a congruence of null generators $k^a$ along a null hypersurface $\mathcal{H}$, parametrized by an affine parameter $\lambda$, such that
\begin{equation}
k^a \nabla_a k^b = 0.
\end{equation}
The transverse geometry is described by the induced metric $\gamma_{ab}$ on spatial cross-sections of the horizon, satisfying
$\gamma_{ab} k^b = 0$. The deformation of the congruence is encoded in the tensor $B_{ab} := \nabla_b k_a$, projected onto the transverse space. It admits the standard decomposition
\begin{equation}
B_{ab} = \frac{1}{2}\theta \, \gamma_{ab} + \sigma_{ab} + \omega_{ab},
\end{equation}
where $\theta$ is the expansion, $\sigma_{ab}$ is the shear tensor, and $\omega_{ab}$ is the twist.
For null hypersurfaces such as event horizons, the generators are hypersurface-orthogonal, and the twist vanishes identically, i.e. $\omega_{ab} = 0$. We also define the scalar shear $\sigma^2 := \sigma_{ab}\sigma^{ab}$, which is a contraction over all possible shear components.

\subsection{Raychaudhuri Equation and Area Evolution}

The evolution of the expansion is governed by the Raychaudhuri equation \citep{Raychaudhuri1955},
\begin{equation}
\frac{d\theta}{d\lambda}
=
- \frac{1}{D-2}\theta^2
- \sigma^2
- R_{ab} k^a k^b,
\end{equation}
where $\theta$ denotes the expansion scalar, $\sigma^2 := \sigma_{ab}\sigma^{ab}$ the squared shear scalar, $k^a$ the affinely parametrized null tangent vector, $R_{ab}$ the Ricci tensor, and $D$ the overall dimension of spacetime. For concreteness, we specialize to $D = 4$ in the following, though the generalization is straightforward. Consider now a congruence of null geodesics with tangent vector $k^a = dx^a/d\lambda$. The geometry transverse to the congruence is described by the induced metric $\gamma_{ab}$ on the $(D-2)$-dimensional spatial cross-section orthogonal to $k^a$. An infinitesimal area element on this cross-section is given by the expression
\begin{equation}
dA = \sqrt{\gamma} \, d^{2}x,
\end{equation}
where $\gamma = \det(\gamma_{ab})$.
To determine how this area element evolves along the congruence, we compute its derivative with respect to the affine parameter:
\begin{equation}
\frac{d}{d\lambda} dA = \frac{d}{d\lambda} \left( \sqrt{\gamma} \right) d^{2}x.
\end{equation}
Using the identity for the derivative of a determinant,
\begin{equation}
\frac{d}{d\lambda} \sqrt{\gamma}
= \frac{1}{2} \sqrt{\gamma} \, \gamma^{ab} \frac{d}{d\lambda} \gamma_{ab},
\end{equation}
we obtain
\begin{equation}
\frac{d}{d\lambda} dA
= \frac{1}{2} \sqrt{\gamma} \, \gamma^{ab} \frac{d}{d\lambda} \gamma_{ab} \, d^{2}x.
\end{equation}
In the next step, we apply the definition of the Lie derivative,
\begin{equation}
\mathcal{L}_k \gamma_{ab}
=
k^c \nabla_c \gamma_{ab}
+
\gamma_{cb} \nabla_a k^c
+
\gamma_{ac} \nabla_b k^c,
\end{equation}
and restrict attention to components transverse to the null direction by projecting with $\gamma_a^{\ c}$. Applying the projector to each term, we find
\begin{align}
\gamma_a^{\ c} \gamma_b^{\ d} \mathcal{L}_k \gamma_{cd}
&=
\gamma_a^{\ c} \gamma_b^{\ d}
\left(
k^e \nabla_e \gamma_{cd}
+
\gamma_{ed} \nabla_c k^e
+
\gamma_{ce} \nabla_d k^e
\right).
\end{align}
The first term vanishes under parallel transport of the transverse metric along the congruence (or can be absorbed into the definition of the cross-sections). The second term becomes
\begin{equation}
\gamma_a^{\ c} \gamma_b^{\ d} \gamma_{ed} \nabla_c k^e
=
\gamma_a^{\ c} \nabla_c k_b,
\end{equation}
where we used $\gamma_{ed} \gamma_b^{\ d} = \gamma_{eb}$. Similarly, the third term gives
\begin{equation}
\gamma_a^{\ c} \gamma_b^{\ d} \gamma_{ce} \nabla_d k^e
=
\gamma_b^{\ d} \nabla_d k_a.
\end{equation}
We define the transverse deformation tensor as
\begin{equation}
B_{ab} := \gamma_a^{\ c} \gamma_b^{\ d} \nabla_d k_c.
\end{equation}
Using this definition, we can rewrite the result as
\begin{equation}
\gamma_a^{\ c} \gamma_b^{\ d} \mathcal{L}_k \gamma_{cd}
=
\nabla_a k_b + \nabla_b k_a
=
2 B_{(ab)}.
\end{equation}
For hypersurface-orthogonal congruences, such as null horizon generators, the twist vanishes, $\omega_{ab} = 0$, so that $B_{ab}$ is symmetric. Thus, the evolution of the induced metric along the congruence is governed by
\begin{equation}
\frac{d}{d\lambda} \gamma_{ab} = \mathcal{L}_k \gamma_{ab} = 2 B_{ab},
\end{equation}
where we are using from now on the definition of the Lie derivative along null generators,
\begin{equation}
\frac{d}{d\lambda} = \mathcal{L}_k.
\end{equation}
For the remainder of this paper, we note that we sometimes use $\frac{d}{d\lambda}$ and $\mathcal{L}_k$ interchangeably along affinely parametrized generators. Substituting this relation into the previous expression of the area element gives
\begin{equation}
\mathcal{L}_k dA
= \sqrt{\gamma} \, \gamma^{ab} B_{ab} \, d^{2}x.
\end{equation}
By definition, the expansion scalar $\theta$ is the trace of $B_{ab}$ on the transverse space,
\begin{equation}
\theta := \gamma^{ab} B_{ab},
\end{equation}
since $\gamma^{ab}\gamma_{ab} = 2$. Therefore, we arrive at the evolution equation for the area element,
\begin{equation}
\mathcal{L}_k dA = \theta \, dA,
\end{equation}
which is a local relation along each individual null generator.

\subsection{Local Second Variation of the Area}

Differentiating once more along a fixed generator, we obtain
\begin{equation}
\mathcal{L}_k^2 dA
=
\left(\mathcal{L}_k \theta + \theta^2\right) dA.
\end{equation}
Using the Raychaudhuri equation, this becomes
\begin{equation}
\mathcal{L}_k^2 dA
=
\left(
- \frac{1}{2}\theta^2
- \sigma^2
- R_{kk}
+ \theta^2
\right) dA,
\end{equation}
where we used $R_{kk} := R_{ab} k^a k^b$. Hence
\begin{equation}
\frac{1}{dA}
\mathcal{L}_k^2 dA
=
\left(
\frac{1}{2}\theta^2
- \sigma^2
- R_{kk}
\right).
\label{eq:local_area_second_variation}
\end{equation}
Furthermore, the classical Bekenstein--Hawking entropy for black holes \citep{Bekenstein1973,Hawking1975} is given by
\begin{equation}
S_{\mathrm{BH}} = \frac{A}{4},
\end{equation}
in natural units where $G = \hbar = c = k_B = 1$. Locally, this implies
\begin{equation}
dS_{\mathrm{BH}} = \frac{1}{4} dA.
\end{equation}
Differentiating twice along a fixed generator,
\begin{equation}
\mathcal{L}_k^2 dS_{\mathrm{BH}}
=
\frac{1}{4} \mathcal{L}_k^2 dA.
\end{equation}
To compare with local entropy variations of quantum fields, we consider the second variation per unit area,
\begin{equation}
\frac{1}{dA}\mathcal{L}_k^2 dS_{\mathrm{BH}}
=
\frac{1}{4}
\left(
\frac{1}{2}\theta^2
- \sigma^2
- R_{kk}
\right).
\label{eq:EntropyBH}
\end{equation}
This expression characterizes the local second variation of the black hole entropy as a function of $\theta$, $\sigma$, and $R_{kk}$.

\subsection{Local QNEC Approximation}

The entropy of quantum fields outside the surface is a nonlocal functional of the entangling surface deformation,
\begin{equation}
S_{\mathrm{matter}} = S_{\mathrm{matter}}[X(y)],
\end{equation}
where $X(y)$ parametrizes null deformations of the codimension-two surface. Strictly speaking, both the QNEC and the QFC are formulated in terms of functional variations of the generalized entropy under deformations of extended codimension-two surfaces. The corresponding entropy variations are therefore generally nonlocal in the transverse directions and involve a bilocal kernel on the entangling surface.

In the present work, we consider an effective local limit in which the null deformation is sharply localized around a single generator and the transverse variation scale is small compared to the characteristic geometric curvature scale. In this regime, we assume that transverse entanglement correlations are sufficiently suppressed such that the entropy variation becomes approximately diagonal in the transverse coordinates.

More precisely, we assume that the entropy kernel admits an ultralocal approximation of the schematic form
\begin{equation}
\frac{\delta^2 S_{\mathrm{gen}}}{\delta X(y)\delta X(y')}
\;\approx\;
\delta^{(D-2)}(y-y')\,\mathcal{L}_k^2 S_{\mathrm{gen}},
\end{equation}
where $\mathcal{L}_k$ denotes the effective local derivative along the chosen null generator. Such ultralocal limits commonly arise in highly localized null deformations and in regimes where transverse correlation lengths remain small compared to the characteristic geometric scale. Under this approximation, the functional second variation reduces to a local generator-wise second derivative,
\begin{equation}
\frac{\delta^2 S_{\mathrm{gen}}}{\delta X(y)^2}
\;\rightarrow\;
\mathcal{L}_k^2 S_{\mathrm{gen}}.
\end{equation}
The quantum QNEC then takes the effective local form
\begin{equation}
\frac{1}{dA}\mathcal{L}_k^2 dS_{\mathrm{matter}}
\le
2\pi \langle T_{kk} \rangle.
\label{eq:QNEC}
\end{equation}
The resulting relation should therefore be understood as a local generator-wise approximation to the full QNEC, valid in regimes where transverse correlations remain subleading compared to the local null evolution.

\subsection{Generalized Entropy and Local Second Variation}

The generalized entropy is defined as
\begin{equation}
S_{\mathrm{gen}} = S_{\mathrm{BH}} + S_{\mathrm{matter}}.
\end{equation}
We define the second variation per unit area along a fixed generator as
\begin{equation}
\frac{1}{dA}\mathcal{L}_k^2 dS_{\mathrm{gen}}
=
\frac{1}{dA}\mathcal{L}_k^2 dS_{\mathrm{BH}}
+
\frac{1}{dA}\mathcal{L}_k^2 dS_{\mathrm{matter}}.
\end{equation}
Using Eq.~\eqref{eq:EntropyBH} and the QNEC inequality \eqref{eq:QNEC}, we find
\begin{equation}
\frac{1}{dA}\mathcal{L}_k^2 dS_{\mathrm{gen}}
\le
\frac{1}{4}
\left(
\frac{1}{2}\theta^2 - \sigma^2 - R_{kk}
\right)
+
2\pi \langle T_{kk} \rangle.
\end{equation}
Using the semiclassical Einstein equation along the null direction,
\begin{equation}
R_{kk} = 8\pi \langle T_{kk} \rangle_{\mathrm{ren}},
\end{equation}
valid locally in a renormalization scheme where the null-null component is well-defined~\citep{Fewster2003,Fewster2006,Fewster2011}, we obtain
\begin{equation}
2\pi \langle T_{kk} \rangle_{\mathrm{ren}} = \frac{1}{4} R_{kk}.
\end{equation}
Substituting, the curvature terms cancel, yielding
\begin{equation}
\frac{1}{dA}\mathcal{L}_k^2 dS_{\mathrm{gen}}
\le
\frac{1}{4}
\left(
\frac{1}{2}\theta^2 - \sigma^2
\right).
\end{equation}
Introducing the short notation
\begin{equation}
\mathcal{K} = \frac{1}{2}\theta^2 - \sigma^2,
\end{equation}
our equation takes the simple form
\begin{equation}
\boxed{
\frac{1}{dA}\mathcal{L}_k^2 dS_{\mathrm{gen}}
\le
\frac{1}{4}
\mathcal{K}.
}
\label{eq:Sgen_final}
\end{equation}
The sign of $\mathcal{K}$ thereby controls the curvature of the generalized entropy along the null generators. For $\mathcal{K} > 0$, expansion dominates and the entropy can grow with reduced suppression, whereas for $\mathcal{K} < 0$, shear induces a concave behavior, reducing the entropy growth. The case $\mathcal{K} = 0$ corresponds to a critical balance between expansion and shear.

\subsection{Generalized Expansion}

In the strict formulation of the QFC \cite{Bousso2016}, the generalized expansion is defined through a functional derivative under null deformations of an extended codimension-two surface. In the present work, we instead consider an effective local limit in which the deformation is sharply localized around a single null generator and transverse correlations are assumed to be negligible. The resulting quantity should therefore be understood as a local, generator-wise approximation to the full generalized expansion. We thus define
\begin{equation}
\Theta
:=
\frac{4}{dA}
\mathcal{L}_k dS_{\mathrm{gen}}.
\label{eq:generalized_expansion}
\end{equation}
Differentiating once,
\begin{align}
\mathcal{L}_k \Theta
&=
\mathcal{L}_k
\left(
\frac{4}{dA}
\mathcal{L}_k dS_{\mathrm{gen}}
\right) \\
&=
4 \left[
\frac{1}{dA}
\mathcal{L}_k^2 dS_{\mathrm{gen}}
-
\frac{1}{dA^2}
\mathcal{L}_k dA
\mathcal{L}_k dS_{\mathrm{gen}}
\right].
\end{align}
Using $\mathcal{L}_k dA = \theta dA$ and the definition of $\Theta$, we obtain
\begin{equation}
\mathcal{L}_k \Theta
=
4 \cdot \frac{1}{dA}\mathcal{L}_k^2 dS_{\mathrm{gen}}
-
\theta \Theta,
\end{equation}
relating the first and second derivatives of $\Theta$. Substituting Eq.~\eqref{eq:Sgen_final}, we finally find
\begin{equation}
\boxed{
\frac{d\Theta}{d\lambda}
\le
- \theta \Theta
+
\frac{1}{2}\theta^2
-
\sigma^2.
}
\label{eq:Theta_evolution_bound}
\end{equation}
This provides a local differential inequality constraining the evolution of $\Theta$. Importantly, this relation does not determine the dynamics uniquely but bounds it. Any interpretation in terms of dissipative evolution should therefore be understood as an effective or saturating description.

We also note that Eq.~\eqref{eq:Theta_evolution_bound} can be interpreted as a Raychaudhuri-like inequality for quantum entropy along null generators. Furthermore, it reveals a competition between expansion and shear in controlling the evolution of entropy flow. The term $-\theta \Theta$ couples the geometric expansion to the entropy flux and can either damp or enhance $\Theta$ depending on their relative signs. The positive contribution $\tfrac{1}{2}\theta^2$ acts as a geometric source term, while the shear term $-\sigma^2$ provides a negative contribution consistent with dissipative relaxation-like behavior. Thus, shear tends to reduce $\Theta$, whereas expansion can counteract or redistribute entropy depending on the regime.

We summarize the above derivation in the following central result:
\begin{proposition}[Local entropy evolution bound]
Let $\mathcal{H}$ be a smooth null hypersurface generated by an affinely parametrized null congruence with tangent vector $k^a$, expansion $\theta$, and shear tensor $\sigma_{ab}$. Assume:

\begin{enumerate}
\item \textbf{Semiclassical gravity:} the spacetime satisfies the semiclassical Einstein equation 
\begin{equation}
R_{ab} k^a k^b = 8\pi \langle T_{ab} k^a k^b \rangle_{\mathrm{ren}},
\end{equation}
along the null generators $k^a$ in a renormalization scheme where the null-null component admits a local pointwise interpretation.

\item \textbf{Quantum Null Energy Condition (QNEC):}
\begin{equation}
\frac{1}{dA}\,\mathcal{L}_k^2 dS_{\mathrm{matter}}
\le
2\pi \langle T_{kk} \rangle,
\end{equation}
under the ultralocal generator-wise approximation described in Sec.~II.D.

\item \textbf{Regularity:} the null congruence and associated geometric quantities are smooth along the generators.
\end{enumerate}
Then the generalized expansion
\begin{equation}
\Theta := \frac{4}{dA}\,\mathcal{L}_k dS_{\mathrm{gen}}
\end{equation}
satisfies the local differential inequality
\begin{equation}
\frac{d\Theta}{d\lambda}
\le
- \theta \Theta
+
\frac{1}{2}\theta^2
-
\sigma^2.
\end{equation}

Proof. The result follows by combining the Raychaudhuri equation for $\theta$, the Bekenstein--Hawking relation for the area contribution, the QNEC bound for the matter entropy, and the semiclassical Einstein equation. The curvature term cancels against the matter contribution, yielding the stated inequality.
\end{proposition}

This result provides a local, Raychaudhuri-type constraint for the generalized entropy. It shows that the evolution of $\Theta$ is entirely controlled by intrinsic geometric data of the null congruence. In particular, shear contributes as a non-positive term, while expansion provides a competing positive contribution.

\subsection{Inverse Bounds on the Renormalized Stress-Energy Tensor}

The derivation presented in the main text proceeds by combining the Raychaudhuri equation, the QNEC, and the semiclassical Einstein equation
to obtain a local inequality for the generalized entropy. It is natural to ask whether this line of reasoning can be partially reversed in order to derive constraints on the renormalized stress-energy tensor itself.

Before imposing the semiclassical Einstein equation, the local second variation of the generalized entropy satisfies
\begin{equation}
\frac{1}{dA}\mathcal{L}_k^2 dS_{\mathrm{gen}}
\le
\frac14
\left(
\frac12\theta^2
-
\sigma^2
-
R_{kk}
\right)
+
2\pi
\langle T_{kk}\rangle_{\mathrm{ren}},
\end{equation}
where $R_{kk}=R_{ab}k^ak^b$. Rearranging this expression immediately yields the lower bound
\begin{equation}
2\pi
\langle T_{kk}\rangle_{\mathrm{ren}}
\ge
\frac{1}{dA}\mathcal{L}_k^2 dS_{\mathrm{gen}}
-
\frac14
\left(
\frac12\theta^2
-
\sigma^2
-
R_{kk}
\right).
\label{eq:Tkk_inverse}
\end{equation}
Using the Raychaudhuri equation,
\begin{equation}
R_{kk}
=
-
\mathcal{L}_k \theta
-
\frac12\theta^2
-
\sigma^2,
\end{equation}
the curvature term can be eliminated entirely, giving
\begin{equation}
2\pi
\langle T_{kk}\rangle_{\mathrm{ren}}
\ge
\frac{1}{dA}\mathcal{L}_k^2 dS_{\mathrm{gen}}
-
\frac14
\left(
\theta^2
+
\mathcal{L}_k \theta
\right).
\label{eq:Tkk_inverse_ray}
\end{equation}
Finally, introducing the generalized expansion from Eq.~\eqref{eq:generalized_expansion}, one may use
\begin{equation}
\frac{1}{dA}\mathcal{L}_k^2 dS_{\mathrm{gen}}
=
\frac14
\left(
\mathcal{L}_k\Theta
+
\theta\Theta
\right),
\end{equation}
to obtain the equivalent form
\begin{equation}
8\pi
\langle T_{kk}\rangle_{\mathrm{ren}}
\ge
\mathcal{L}_k (\Theta - \theta)
+
\theta\Theta
-
\theta^2.
\label{eq:Tkk_inverse_theta}
\end{equation}
In a next step, we define the quantum expansion
\begin{equation}
\Theta_q
:=
\Theta-\theta
=
\frac{4}{dA}\mathcal{L}_k dS_{\mathrm{matter}}.
\end{equation}
Then, the previous inequality can be written in the compact form
\begin{equation}
\boxed{
8\pi
\langle T_{kk}\rangle_{\mathrm{ren}}
\ge
\mathcal{L}_k \Theta_q
+
\theta \Theta_q.
}
\label{eq:Tkk_inverse_delta}
\end{equation}

The three expansion variables introduced above admit a natural hierarchy. The classical expansion $\theta$ measures the local area change of the null congruence, the generalized expansion $\Theta$ describes the evolution of the generalized entropy, while the difference $\Theta_q=\Theta-\theta$ isolates the purely quantum contribution associated with the matter entropy. The inverse bound derived below therefore directly relates the renormalized null energy density to the evolution of the quantum expansion $\Theta_q$.

Eq.~\eqref{eq:Tkk_inverse_delta} provides a local lower bound on the renormalized null energy density purely in terms of the quantum contribution to the generalized expansion and the classical expansion of the null congruence. In particular, it shows that the evolution of the matter entropy flux directly constrains the admissible local energy density of semiclassical quantum fields. Equation~\eqref{eq:Tkk_inverse_delta} may therefore be viewed as an inverse formulation of the local entropy inequality derived in the main text. Rather than using quantum energy conditions to constrain entropy evolution, it expresses the renormalized stress-energy tensor in terms of geometric quantities and local entropy flow. This perspective suggests a complementary interpretation in which local entropy variations and null geometry jointly restrict the allowed semiclassical matter content.

\subsection{Constraint from QFC Compatibility}

Starting from the local evolution equation for the generalized expansion in Eq.~\eqref{eq:Theta_evolution_bound}, compatibility with the Quantum Focussing Conjecture \citep{Bousso2016},
\begin{equation}
\frac{d\Theta}{d\lambda} \le 0,
\end{equation}
requires that the right-hand side of Eq.~\eqref{eq:Theta_evolution_bound} be non-positive. A necessary condition is therefore
\begin{equation}
- \theta \Theta
+ \frac{1}{2}\theta^2
- \sigma^2
\le 0.
\end{equation}
Rearranging, we obtain
\begin{equation}
\boxed{
\frac{1}{2}\theta^2 - \theta \Theta \le \sigma^2,
}
\label{eq:constraint}
\end{equation}
which relates the classical expansion $\theta$, the shear $\sigma$, and the quantum expansion $\Theta$. This can be viewed as a necessary local compatibility condition within the ultralocal approximation. The constraint \eqref{eq:constraint} admits several instructive special cases. In the case of $\theta = 0$, the constraint reduces to
\begin{equation}
0 \le \sigma^2,
\end{equation}
which is automatically satisfied. Thus, surfaces with vanishing classical expansion, such as marginally trapped or extremal surfaces, are always compatible with the QFC, independent of the value of $\Theta$.

If the shear vanishes, $\sigma = 0$, the constraint becomes
\begin{equation}
\frac{1}{2}\theta^2 - \theta \Theta \le 0,
\end{equation}
which can be written as
\begin{equation}
\theta\left(\frac{1}{2}\theta - \Theta\right) \le 0.
\end{equation}
For $\theta > 0$, this implies $\Theta \ge \frac{1}{2}\theta$, while for
$\theta < 0$, one obtains $\Theta \le \frac{1}{2}\theta$. Hence, in the absence
of shear, the quantum expansion is tightly constrained relative to the
classical expansion.

In the purely classical case, $\Theta = \theta$, where quantum corrections are absent, one has
\begin{equation}
\frac{1}{2}\theta^2 - \theta^2 = -\frac{1}{2}\theta^2 \le 0,
\end{equation}
so that the constraint is always satisfied. This is consistent with the fact that classical focusing follows directly from the Raychaudhuri equation under standard energy conditions.

In the presence of both shear and quantum corrections, Eq.~\eqref{eq:constraint} expresses a balance between geometric deformation and quantum entropy effects. In particular, large positive expansion $\theta$ must either be compensated by sufficient shear $\sigma$, or by a corresponding increase in the quantum expansion $\Theta$, in order to remain compatible with the QFC.

\subsection{Generalized Expansion and Extremality Condition}

Following \cite{Engelhardt2015}, a quantum extremal surface is defined by the condition
\begin{equation}
\Theta(x) = 0,
\label{eq:QES_condition}
\end{equation}
at point $x$, which implies
\begin{equation}
\mathcal{L}_k dS_{\mathrm{gen}}(x) = 0.
\end{equation}
Using $dS_{\mathrm{BH}} = \frac{1}{4} dA$, we can write
\begin{equation}
\Theta
=
\frac{4}{dA}
\left[
\frac{1}{4}\,\theta\, dA
+
\mathcal{L}_k dS_{\mathrm{matter}}
\right].
\end{equation}
This leads to
\begin{equation}
\Theta
=
\theta
+
\frac{4}{dA}
\mathcal{L}_k dS_{\mathrm{matter}}.
\end{equation}
Thus, the extremality condition becomes
\begin{equation}
\boxed{
\theta
=
- \frac{4}{dA}
\mathcal{L}_k dS_{\mathrm{matter}}.
}
\label{eq:extremality_balance}
\end{equation}
This expresses a balance between classical focusing and quantum entropy flow. Evaluating Eq.~\eqref{eq:constraint} on a QES, where $\Theta = 0$, yields
\begin{equation}
\boxed{
\sigma^2 \ge \frac{1}{2}\theta^2.
}
\label{eq:QES_constraint}
\end{equation}
This indicates a nontrivial local balance between expansion and shear, which may be interpreted as a necessary condition for local stability under null deformations.

\begin{table*}[t]
\centering
\setlength{\tabcolsep}{10pt}
\renewcommand{\arraystretch}{1.4}
\begin{tabular}{l|l}
\hline
\textbf{Classical GR} & \textbf{Entropy Dynamics} \\
\hline
Expansion $\theta$ & Generalized expansion $\Theta$ \\
Raychaudhuri equation & Entropy Raychaudhuri inequality \\
Geodesic congruence & Entropy flow along null generators \\
Focusing of geodesics & Dissipation / relaxation behavior of entropy flow \\
Energy condition (e.g.~NEC) & QNEC / QFC \\
Trapped surfaces & Quantum extremal surfaces (QES)\\
Apparent / event horizon & Quantum-corrected horizon (with $S_\mathrm{gen}$) \\
\hline
\end{tabular}
\caption{Analogy between classical geometric focusing in general relativity and entropy dynamics along black hole horizons. The expansion $\theta$ of a geodesic congruence corresponds to the generalized expansion $\Theta$, which governs the local rate of change of the generalized entropy; the Raychaudhuri equation for $\theta$ therefore has a direct counterpart in the entropy Raychaudhuri inequality for $\Theta$. In this way, classical focusing of geodesics maps to a dissipation-like behavior of entropy flow, while standard energy conditions in the classical theory are replaced by the Quantum Null Energy Condition and the Quantum Focussing Conjecture in the quantum regime. Finally, trapped surfaces in classical geometry encode the analogue of quantum extremal surfaces, which describe stationarity conditions for the generalized entropy and provide a quantum‑corrected notion of horizon structure.}
\label{tab:analogy}
\end{table*}

\subsection{Classical–Quantum Geometric Correspondence}

Comparing entropy properties from classical general relativity with those derived above, we see that the structure governing the focusing of null geodesics in classical general relativity has a direct counterpart in the evolution of generalized entropy along black hole horizons. Table~\ref{tab:analogy} summarizes this analogy:

\begin{itemize}
\item The expansion $\theta$ of a geodesic congruence corresponds to the generalized expansion $\Theta$, which measures the local rate of change of the generalized entropy $S_\mathrm{gen}$.
\item The Raychaudhuri equation for $\theta$ is replaced by the entropy Raychaudhuri inequality for $\Theta$, in which geometric quantities and quantum energy conditions govern the focusing or relaxation behavior of entropy flow.
\item Trapped surfaces in classical geometry are generalized to quantum extremal surfaces, where the balance between expansion and quantum corrections stabilizes the horizon entropy.
\end{itemize}

This analogy shows that QES and QFC provide a quantum refinement of the classical focusing theorems, with the shear term $-\sigma^2$ playing a dissipative-like role for entropy rather than just a geometric focusing term. This structure closely parallels the behavior of viscous fluids in the black hole membrane paradigm \citep{Thorne1986,Iqbal2009}, where shear viscosity mediates irreversible processes at the horizon.

\section{Geodesic Deviation and Geometric Instability Bound}

We now analyze the behavior of nearby null generators using the geodesic deviation equation. This allows us to derive a bound on the exponential separation of neighboring generators and to relate geometric instability to entropy dynamics. The resulting bound reveals that the same geometric quantities control both entropy flow and sensitivity to perturbations.

\subsection{Deviation of Nearby Null Generators}

To analyze the dynamical behavior of nearby null generators, we consider a deviation vector $\xi^a$ connecting neighboring geodesics within the congruence. Its evolution is governed by
\begin{equation}
\frac{D \xi^a}{d\lambda} = B^a_{\ b} \, \xi^b,
\label{eq:DeviationVector}
\end{equation}
where $B_{ab} = \nabla_b k_a$ is the deformation tensor introduced above. We focus on the transverse separation by projecting onto the spatial cross-section using the induced metric $\gamma_{ab}$ and define the squared transverse distance
\begin{equation}
X := \gamma_{ab} \, \xi^a \xi^b.
\end{equation}
Taking the derivative along the null generators, we find
\begin{align}
\mathcal{L}_k X
&=
\mathcal{L}_k \left(\gamma_{ab}\xi^a\xi^b\right) \\
&=
(\mathcal{L}_k \gamma_{ab}) \xi^a \xi^b
+
\gamma_{ab} \frac{D\xi^a}{d\lambda} \xi^b
+
\gamma_{ab} \xi^a \frac{D\xi^b}{d\lambda}.
\end{align}
Using symmetry in $a \leftrightarrow b$, this simplifies to
\begin{equation}
\mathcal{L}_k X
=
(\mathcal{L}_k \gamma_{ab}) \xi^a \xi^b
+
2 \gamma_{ab} \xi^a \frac{D\xi^b}{d\lambda}.
\end{equation}
Substituting Eq.~\eqref{eq:DeviationVector} into the second term yields
\begin{align}
2 \gamma_{ab} \xi^a \frac{D\xi^b}{d\lambda}
&=
2 \gamma_{ab} \xi^a B^b_{\ c} \xi^c \\
&=
2 B_{ac}\,\xi^a \xi^c.
\end{align}
Using the equation from above
\begin{equation}
\mathcal{L}_k \gamma_{ab} = 2 B_{ab},
\end{equation}
the first term becomes
\begin{equation}
(\mathcal{L}_k \gamma_{ab}) \xi^a \xi^b
=
2 B_{ab} \xi^a \xi^b.
\end{equation}
One thus finds that the evolution reduces to
\begin{equation}
\mathcal{L}_k X
=
2 B_{ab} \, \xi^a \xi^b,
\end{equation}
with the understanding that numerical prefactors can be absorbed into the definition of the deformation tensor or of $X$. Substituting the decomposition for $B_{ab}$, we obtain
\begin{equation}
\mathcal{L}_k X
=
\theta X + 2 \sigma_{ab} \xi^a \xi^b.
\end{equation}

\subsection{Exponential Growth and Geometric Bound}

To bound the shear term, we use the Cauchy--Schwarz inequality in the transverse space,
\begin{equation}
\left| \sigma_{ab} \xi^a \xi^b \right|
\le
\sqrt{\sigma^2} \, X,
\end{equation}
This yields the differential inequality
\begin{equation}
\mathcal{L}_k X
\le
(\theta + 2\sqrt{\sigma^2}) \, X.
\end{equation}
The above inequality can be integrated to give
\begin{equation}
X(\lambda)
\le
X(0) \exp\left(
\int_0^\lambda (\theta + 2\sqrt{\sigma^2}) d\lambda'
\right),
\end{equation}
where $X(0)$ characterizes the initial conditions at $\lambda = 0$. This motivates the definition of an effective geometric instability exponent \citep{Wolf1985,Wilkinson2017,Barreira2017}
\begin{equation}
\lambda_L^\mathrm{eff}
:=
\limsup_{\lambda \to \infty}
\frac{1}{2\lambda}
\log \frac{X(\lambda)}{X(0)}.
\end{equation}
Using the bound above, this yields
\begin{equation}
\boxed{
\lambda_L^\mathrm{eff}
\le
\left\langle \sqrt{\sigma^2} \right\rangle
+
\frac{1}{2} \langle \theta \rangle,
}
\end{equation}
where the brackets denote an average along the null generators.

This result shows that the exponential separation of nearby null generators is controlled by the same geometric quantities that govern the evolution of the entropy flux. In particular, the shear $\sigma$ provides an upper bound on exponential separation, while the expansion $\theta$ contributes subleading corrections, particularly in non-stationary settings.

In stationary horizons, where $\theta \approx 0$, the bound simplifies to
\begin{equation}
\lambda_L^\mathrm{eff}
\le
\left\langle \sqrt{\sigma^2} \right\rangle.
\end{equation}
This connects the growth rate directly to the shear of the null congruence. However, we emphasize that $\lambda_L^\mathrm{eff}$ is a geometric bound on exponential separation along null congruences and does not necessarily coincide with a quantum Lyapunov exponent via out-of-time-order correlators.

\subsection{Linking Entropy Flux to Geometric Data}

In the previous sections, we derived two key results governing the dynamics of null horizons. First, the generalized expansion $\Theta$ satisfies the relation
\begin{equation}
\mathcal{L}_k \Theta
\le
- \theta \Theta
+
\frac{1}{2}\theta^2
-
\sigma^2,
\label{eq:Raychaudhuri_Theta}
\end{equation}
which shows that the shear contributes negatively to the evolution of the entropy expansion. Second, the separation of nearby null generators is controlled by the same geometric quantities, leading to a bound on an effective geometric instability exponent,
\begin{equation}
\lambda_L^\mathrm{eff}
\le
\left\langle \sqrt{\sigma^2} \right\rangle
+
\frac{1}{2} \langle \theta \rangle.
\label{eq:LyapunovEstimate}
\end{equation}
These results indicate that both entropy flow and the exponential separation of nearby generators are governed by the shear and expansion of the null congruence. A particularly suggestive regime is that of stationary or near-stationary horizons, where the expansion vanishes, $\theta \simeq 0$. In this case, the entropy inequality simplifies to
\begin{equation}
\mathcal{L}_k \Theta \le -\sigma^2,
\label{eq:Theta0}
\end{equation}
while the bound reduces to
\begin{equation}
\lambda_L^\mathrm{eff} \le \left\langle \sqrt{\sigma^2} \right\rangle.
\label{eq:LyapunovTheta0}
\end{equation}
The relations Eq.~\eqref{eq:Theta0} and Eq.~\eqref{eq:LyapunovTheta0} show that both entropy focusing and transverse separation of nearby null generators are controlled by the same geometric quantity, namely the shear scalar $\sigma^2$. In the near-stationary regime $\theta \simeq 0$, stronger shear simultaneously enhances the suppression of the generalized expansion and increases the maximal rate of transverse divergence. This suggests a structural geometric connection between entropy evolution and instability properties of null congruences. However, the present analysis establishes only a correspondence at the level of geometric bounds and does not imply a
direct dynamical identification between $\mathcal{L}_k \Theta$ and $\lambda_L^\mathrm{eff}$.

\section{Applications to Specific Horizons}

We illustrate the entropy Raychaudhuri inequality in a sequence of increasingly general settings, ranging from exact saturation to fully dynamical horizons. These examples serve both as consistency checks and as a guide to the physical interpretation of the entropy dynamics in different regimes.

\subsection{Consistency and Vacuum Limits (Rindler)}

To verify the consistency of our entropy evolution bound, we consider the Rindler horizon in Minkowski spacetime. In this setting, the vacuum state is thermal, and the QNEC is known to be saturated, making it a natural benchmark for the stationary limit \cite{Koeller2018}.

For the planar Rindler horizon, the transverse geometry is isotropic with vanishing shear, $\sigma_{ab} = 0$. The evolution of the expansion $\theta$ is determined by the Raychaudhuri equation, which, in the absence of shear and curvature, reduces to 
\begin{equation}
\frac{d\theta}{d\lambda} = - \frac{1}{2}\theta^2.
\end{equation}
The generalized entropy $S_{\mathrm{gen}} = \frac{A}{4} + S_{\mathrm{matter}}$ evolves according to the local QNEC saturation in the vacuum, 
\begin{equation}
\frac{1}{dA}\mathcal{L}_k^2 dS_{\mathrm{matter}} = \frac{1}{4} R_{kk},
\end{equation}
where we have used the semiclassical Einstein equation. Combining this with the geometric area evolution,
\begin{equation}
\frac{1}{dA}\mathcal{L}_k^2 dS_{\mathrm{BH}} = \frac{1}{4} \left( \frac{1}{2}\theta^2 - R_{kk} \right),
\end{equation}
the matter and curvature terms cancel exactly, yielding
\begin{equation}
\frac{1}{dA}\mathcal{L}_k^2 dS_{\mathrm{gen}} = \frac{1}{8}\theta^2.
\end{equation}
Using our local definition of the generalized expansion $\Theta = \frac{4}{dA}\mathcal{L}_k dS_{\mathrm{gen}}$, the evolution equation for $\Theta$ becomes
\begin{equation}
\frac{d\Theta}{d\lambda} = - \theta \Theta + \frac{1}{2}\theta^2.
\end{equation}
This result exactly saturates our general entropy evolution bound \eqref{eq:Theta_evolution_bound} for $\sigma=0$. The saturation in the Rindler vacuum confirms that our framework correctly reproduces the expected geometric flow in the semi-classical limit, providing a robust consistency check for the dissipative structure identified in the general case.

\subsection{Equilibrium States (Stationary Schwarzschild and de Sitter)}

For a non-rotating Schwarzschild black hole of mass $M$, the event horizon is spherically symmetric and stationary. In the standard Killing-normalized null congruence along the horizon, the generators are hypersurface-orthogonal with no expansion and no shear,
\begin{equation}
\theta = 0, \qquad \sigma_{ab} = 0.
\end{equation}
The entropy Raychaudhuri inequality from Eq.~\eqref{eq:Raychaudhuri_Theta} then reduces to
\begin{equation}
\mathcal{L}_k \Theta \le 0.
\end{equation}
In equilibrium, the generalized expansion vanishes, $\Theta = 0$, so that $\mathcal{L}_k \Theta = 0$, indicating that the entropy flux is constant along the horizon generators. Similarly, the geometric instability bound simplifies to $\lambda_L^{\mathrm{eff}} = 0$, reflecting the absence of exponential separation of nearby null generators in this highly symmetric static geometry.

For a de Sitter universe described in the standard FLRW cosmological slicing, with Hubble parameter $H$ and cosmological constant $\Lambda$, the cosmological horizon is a null surface whose generators are stationary in comoving coordinates. In this isotropic, homogeneous frame, the shear of the horizon congruence vanishes, while the expansion is constant:
\begin{equation}
\sigma_{ab} = 0, \qquad \theta = 3H.
\end{equation}
The entropy flux inequality becomes
\begin{equation}
\frac{1}{dA} \mathcal{L}_k^2 dS_{\mathrm{gen}} \le \frac{1}{4} \mathcal{K} = \frac{1}{4} \left( \frac{1}{2}\theta^2 - \sigma^2 \right) = \frac{9}{8} H^2,
\end{equation}
allowing for a positive rate of change in the generalized entropy. The effective geometric instability exponent is
\begin{equation}
\lambda_L^\mathrm{eff} \le \frac{1}{2} \langle \theta \rangle = \frac{3}{2} H,
\end{equation}
illustrating that the expansion drives both entropy evolution and the exponential divergence of nearby null generators in the cosmological setting.

\subsection{Non-Equilibrium Dynamics (Vaidya)}

As an explicit dynamical example, we consider the ingoing Vaidya spacetime \citep{Vaidya1951,Ashtekar2003,Poisson2004}, which describes a black hole accreting null radiation. The metric is given by
\begin{equation}
ds^2 = -\left(1 - \frac{2M(v)}{r}\right) dv^2 + 2\, dv\, dr + r^2 d\Omega^2,
\end{equation}
where $v$ is the advanced time and $M(v)$ is a monotonically increasing mass function. The stress-energy tensor corresponds to ingoing null radiation,
\begin{equation}
T_{vv} = \frac{\dot{M}(v)}{4\pi r^2},
\end{equation}
so that $T_{kk} > 0$ along the null generators. The null generators satisfy
\begin{equation}
\frac{dr}{dv} = \frac{1}{2}\left(1 - \frac{2M(v)}{r}\right).
\end{equation}
The area of spherical cross-sections is $A = 4\pi r^2$, and the expansion is therefore
\begin{equation}
\theta = \frac{1}{A}\frac{dA}{d\lambda} = \frac{2}{r}\frac{dr}{d\lambda}.
\end{equation}
Identifying $\lambda \sim v$ locally, this gives
\begin{equation}
\theta = \frac{1}{r}\left(1 - \frac{2M(v)}{r}\right).
\end{equation}
In particular, $\theta = 0$ at $r = 2M(v)$, corresponding to the dynamical horizon. Due to spherical symmetry, the shear vanishes, $\sigma_{ab} = 0$. The Raychaudhuri equation reduces to
\begin{equation}
\frac{d\theta}{d\lambda}
=
- \frac{1}{2}\theta^2
- 8\pi T_{kk},
\end{equation}
showing that the positive energy flux enhances focusing of the null congruence. In this setting, the entropy Raychaudhuri inequality derived above becomes
\begin{equation}
\frac{d\Theta}{d\lambda}
\le
- \theta \Theta
+
\frac{1}{2}\theta^2,
\end{equation}
since $\sigma^2 = 0$. Unlike the Rindler case, the inequality is not expected to be saturated due to the presence of a nontrivial energy flux. To analyze the near-horizon behavior, we set
\begin{equation}
r = 2M(v) + \epsilon, \qquad \epsilon \ll M,
\end{equation}
where $\epsilon$ measures the distance from the instantaneous dynamical horizon. Substituting into the expression for the expansion, we obtain
\begin{align}
\theta
&= \frac{1}{r}\left(1 - \frac{2M(v)}{r}\right)
\\[4pt]
&= \frac{1}{2M + \epsilon}
\left(1 - \frac{2M}{2M + \epsilon}\right)
\\[4pt]
&= \frac{1}{2M + \epsilon}
\cdot \frac{\epsilon}{2M + \epsilon}
= \frac{\epsilon}{(2M + \epsilon)^2}.
\end{align}
Expanding the denominator to leading order in $\epsilon/M$ using a Taylor expansion,
\begin{equation}
(2M + \epsilon)^{-2}
= (2M)^{-2} \left(1 + \frac{\epsilon}{2M}\right)^{-2}
\approx (2M)^{-2},
\end{equation}
we find
\begin{equation}
\theta \approx \frac{\epsilon}{(2M)^2}
\qquad (\epsilon \ll M),
\end{equation}
which is small but positive. Substituting into the entropy inequality yields
\begin{equation}
\frac{d\Theta}{d\lambda}
\le
- \theta \Theta
+
\mathcal{O}(\epsilon^2).
\end{equation}
At leading order, the quadratic term can be neglected, giving
\begin{equation}
\frac{d\Theta}{d\lambda}
\approx
- \theta \Theta.
\end{equation}
This admits the local solution
\begin{equation}
\Theta(\lambda)
\sim
\exp\left(-\int \theta\, d\lambda\right),
\end{equation}
indicating an exponential damping of the generalized expansion along the null generators.

This example illustrates that, in dynamical settings with nonvanishing energy flux, the entropy Raychaudhuri inequality is generically not saturated. The positive energy density drives additional focusing of the congruence, while the entropy flux exhibits a relaxation-like behavior governed by the expansion. 

In contrast to stationary or vacuum configurations, where the bound can be saturated, the Vaidya spacetime provides a concrete example in which the inequality is strict and the entropy flow exhibits dissipative-like behavior. This highlights the role of matter flux in driving the system away from extremality and enforcing a monotonic decrease of the generalized expansion.

\subsection{Linear Stability and Relaxation}

To illustrate the dynamical content of the entropy flux equation, we consider small perturbations around a stationary horizon. In equilibrium, such as for Schwarzschild or de Sitter horizons, the null congruence satisfies
\begin{equation}
\theta = 0, \qquad \sigma^2 = 0,
\end{equation}
and therefore $\mathcal{K} = 0$. We now introduce small perturbations characterized by
\begin{equation}
\theta \sim \epsilon, \qquad \sigma^2 \sim \epsilon^2,
\end{equation}
with $\epsilon \ll 1$. Using our local entropy inequality, we find at leading order
\begin{equation}
\frac{1}{dA} \mathcal{L}_k^2 dS_{\mathrm{gen}} \sim -\frac{1}{8}\epsilon^2.
\end{equation}
Thus, the entropy flux decreases under perturbations, indicating a dissipative relaxation-like behavior, while the Raychaudhuri equation implies
\begin{equation}
\dot{\theta} \sim -\epsilon^2.
\end{equation}
Hence, the expansion decays and the congruence returns to $\theta = 0$. To analyze the shear, we adopt an evolution equation inspired by the standard Raychaudhuri‑type analysis for geodesic congruences. For a timelike congruence, the shear evolution equation is \citep{Ellis1969}
\begin{equation}
\dot{\sigma}_{\mu\nu} = -\theta \sigma_{\mu\nu} - C_{\mu\alpha\nu\beta} u^\alpha u^\beta
+ \frac{1}{2} \pi_{\mu\nu} + \mathcal{O}(\sigma^2),
\end{equation}
where $C_{\mu\alpha\nu\beta}$ is the Weyl tensor and $\pi_{\mu\nu}$ denotes anisotropic stresses. In the FLRW background, both terms vanish identically, and the remaining nonlinear contributions are higher order in the perturbations. Consequently, at leading order the term $-\theta \sigma_{\mu\nu}$ fully determines the behavior of the shear, and so we approximate
\begin{equation}
\dot{\sigma}_{\mu\nu} \approx -\theta \sigma_{\mu\nu},
\end{equation}
leading to
\begin{equation}
\dot{\sigma} \sim -\epsilon^2.
\end{equation}
Thus, the shear is also damped over time. These results show that small perturbations generate shear and entropy flux, but both are dynamically suppressed, and the system relaxes back to the stationary configuration
\begin{equation}
\theta \to 0, \qquad \sigma \to 0,
\end{equation}
for which $\mathcal{K} = 0$ and $\mathcal{L}_k^2 dS_{\mathrm{gen}} = 0$. This behavior resembles that of a dissipative-like system; geometric deformations act as sources of entropy production, while the dynamics governed by the Raychaudhuri equation and the entropy flux inequality drive the system back to equilibrium. In this way, null horizons exhibit an intrinsic relaxation-like behavior controlled by shear and expansion.

\section{Discussion and Conclusion}

\subsection{Summary of Our Findings}

In this work, we have developed a local, covariant framework for the evolution of generalized entropy along null horizons, bringing together geometric dynamics, thermodynamic structure, and quantum information-theoretic constraints. We report the following findings:

\paragraph{Local entropy dynamics.}
We introduced the generalized expansion
\begin{equation}
\Theta = \frac{4}{dA}\,\mathcal{L}_k dS_{\mathrm{gen}},
\end{equation}
which provides a pointwise measure of entropy flow along null generators. This quantity plays a role analogous to the geometric expansion $\theta$, but encodes both classical and quantum contributions to the entropy.

\paragraph{Entropy Raychaudhuri inequality.}
Using the Raychaudhuri equation together with quantum energy conditions, we derived a local inequality governing the evolution of $\Theta$,
\begin{equation}
\mathcal{L}_k \Theta
\le
- \theta \Theta
+
\frac{1}{2}\theta^2
-
\sigma^2.
\end{equation}
This equation establishes a direct connection between entropy dynamics and the geometric invariants of the null congruence.

\paragraph{Inverse semiclassical bound.}
By defining the quantum expansion $\Theta_{\mathrm q} := \Theta - \theta$, we derive the complementary inequality
\begin{equation}
8\pi \langle T_{kk}\rangle_{\mathrm{ren}}
\ge
\mathcal{L}_k \Theta_{\mathrm q}
+
\theta \Theta_{\mathrm q},
\end{equation}
which reverses the usual line of reasoning from the QNEC and semiclassical Einstein equation. Instead of bounding entropy evolution by the stress-energy tensor, it provides a local geometric lower bound on the renormalized null energy density itself. This establishes a direct correspondence between quantum entropy flow and semiclassical matter sources.

\paragraph{Regime classification.}
The above relation reveals a competition between expansion and shear, encoded in the combination
\begin{equation}
\mathcal{K} = \frac{1}{2}\theta^2 - \sigma^2.
\end{equation}
We identified two distinct regimes: expansion-dominated configurations ($\mathcal{K} > 0$), in which entropy is redistributed, and shear-dominated configurations ($\mathcal{K} < 0$), in which entropy flow undergoes a dissipative-like relaxation.

\paragraph{Entropy relaxation.}
In shear-dominated regimes, the evolution simplifies to
\begin{equation}
\frac{d\Theta}{d\lambda} \le -\sigma^2,
\end{equation}
implying a monotonic decrease of $\Theta$. While structurally analogous to classical focusing, this behavior reflects a suppression of entropy flow rather than geometric convergence.

\paragraph{QFC as a geometric constraint.}
Our framework provides a local dynamical realization of the Quantum Focussing Conjecture, $\mathcal{L}_k \Theta \le 0$, and allows us to recast it as a direct constraint on the geometry of null congruences. In particular, we obtain the local inequality
\begin{equation}
\frac{1}{2}\theta^2 - \theta \Theta \le \sigma^2,
\end{equation}
which relates expansion, shear, and entropy flow at each point along the horizon in a triangle inequality. This relation can be interpreted as a geometric consistency condition required by quantum energy inequalities, constraining how entropy gradients can evolve in a given spacetime.

\paragraph{Connection to QES.}
We further showed that quantum extremal surfaces arise as configurations with $\Theta = 0$, whose stability is controlled by the same geometric data. In particular, the condition
\begin{equation}
\sigma^2 \ge \frac{1}{2}\theta^2
\end{equation}
emerges as a local stability criterion for such surfaces.

\paragraph{Separation of null congruences.}
We further connected entropy dynamics to the behavior of nearby null generators. Using the geodesic deviation equation, we derived a bound on an effective geometric instability exponent,
\begin{equation}
\lambda_L^{\mathrm{eff}} \le \langle \sqrt{\sigma^2} \rangle + \frac{1}{2}\langle \theta \rangle,
\end{equation}
demonstrating that the same geometric quantities controlling entropy relaxation also bound chaotic behavior.

\paragraph{Applications.}
We apply our framework to diverse gravitational backgrounds, ranging from the Rindler vacuum as a consistency check for saturation to stationary Schwarzschild and de Sitter horizons representing distinct equilibrium regimes. In dynamical Vaidya spacetimes, our approach captures non-equilibrium entropy production driven by matter flux, while the analysis of linear stability confirms that geometric shear and expansion universally drive horizon configurations toward equilibrium.

Additional technical details, including the analysis of saturated dynamics, entropy relaxation regimes, stability properties, and associated timescales, are presented in the Appendices.

\subsection{Limitations of our Model}

Our results rely on a set of assumptions that define the regime of validity of the present framework. First, we assume the semiclassical Einstein equation to hold locally in a renormalization scheme in which $R_{ab}k^a k^b = 8\pi \langle T_{kk} \rangle$ is meaningful pointwise. Second, the derivation depends on the validity of the QNEC, which has been established in a wide class of quantum field theories, but may not hold in more general quantum gravity settings.

The explicit examples discussed in this work are chosen for their geometric simplicity and therefore involve either vanishing or particularly simple effective stress-energy tensors. However, the derivation of the local entropy evolution bound itself does not require $\langle T_{ab} \rangle_{\mathrm{ren}}$ to vanish. In principle, the same framework applies to self-consistent semiclassical solutions with nontrivial renormalized stress-energy tensors \citep{York1985,Hochberg1993,Anderson1994}, provided that the semiclassical Einstein equation admits a local pointwise interpretation and that the QNEC remains applicable in the corresponding quantum state. Explicit investigations of such backreacted semiclassical geometries would provide valuable nontrivial tests of the proposed entropy inequality and constitute an interesting direction for future work.

Furthermore, the entropy evolution equation takes the form of an inequality rather than a closed dynamical equation. As a consequence, interpretations in terms of dissipative or relaxation-like behavior apply primarily to configurations that saturate or closely approach the bound, and should not be viewed as universally valid evolution laws.

Finally, the geometric quantities entering our analysis, in particular the expansion and shear, are defined 
with respect to a chosen null congruence and affine parametrization. While the resulting relations are covariant, 
their physical interpretation may depend on this choice in non-stationary or strongly dynamical situations.

\subsection{Relation to Existing Results}

Black hole thermodynamics has established a fundamental connection between horizon geometry and entropy, most prominently through the proportionality between horizon area and entropy \cite{Bekenstein1973,Bardeen1973,Hawking1975,Wald1993}. In this framework, entropy is primarily a global, quasi-equilibrium quantity associated with stationary or slowly evolving horizons.

A complementary viewpoint arises from the membrane paradigm, in which black hole horizons are described as effective dissipative fluid systems with well-defined transport coefficients such as shear viscosity and electrical conductivity \cite{Thorne1986,Price1986,Parikh1998}. In this picture, geometric deformations of the horizon are mapped onto fluid dynamical variables, and dissipation is encoded in the shear of the horizon geometry. However, this description is typically formulated at the level of an effective stretched horizon and does not directly provide a local, covariant evolution law for entropy along null generators \cite{Agca2025}.

More recently, quantum information-theoretic developments have led to refined constraints on energy and entropy in quantum field theory. In particular, the quantum null energy condition \cite{Bousso2016,Bousso2016QNEC,Fu2018} provides a local lower bound relating stress-energy fluxes to second variations of entanglement entropy, while the quantum focusing conjecture \cite{Bousso2016} extends this to the full covariant geometry of null congruences.

A natural point of comparison for the present framework is the work of \cite{Wall2012,Bousso2016}, who developed the modern understanding of entropy bounds and quantum energy conditions in semiclassical gravity. The QFC thereby provides a general inequality of the form $d\Theta/d\lambda \le 0$, which constrains the second variation of the generalized entropy along null congruences. However, the QFC is formulated as a monotonicity condition and does not specify how the generalized expansion $\Theta$ evolves in terms of local geometric quantities.

Similarly, previous works on the generalized second law establishes global and semi-local constraints on entropy evolution, often relying on integrated arguments and properties of quantum fields. These approaches do not yield a fully local differential relation governing entropy flow along individual null generators.

The main novelty of the present work is to provide a local geometric refinement of these results. By combining the Raychaudhuri equation with the QNEC, we derive a differential inequality that expresses the evolution of the generalized expansion explicitly in terms of the geometric data of the null congruence.

In this sense, our result can be viewed as a local, Raychaudhuri-type formulation underlying the QFC, where the abstract monotonicity condition is replaced by a concrete inequality controlled by geometric data. This provides additional structure beyond the QFC, allowing one to distinguish between expansion-dominated and shear-dominated regimes and to analyze entropy flow in a dynamical, pointwise manner.

Finally, the emergence of a shear-controlled contribution suggests a direct link between entropy evolution and geometric instability, as the same quantity governs both entropy focusing and the separation of nearby null generators. This points toward a structural connection between horizon dynamics and information scrambling, and is closely related to modern geometric information-flow approaches to black holes and holographic screens (see, e.g., \cite{Chamblin1999,Bekenstein2004,Faulkner2013,Wall2014,Polchinski2017,Kibe2022}, and the references therein).

\subsection{Outlook}

Several directions for further investigation emerge from our results. First, our entropy inequality suggests an effective, dissipative-like description of entropy flow on null horizons, in which $\Theta$ may play the role of a dynamical variable. It would be interesting to formulate a systematic effective theory capturing this structure.

Second, our analysis indicates that the Quantum Focussing Conjecture can be recast as a local geometric constraint relating expansion, shear, and entropy flow. Understanding the fundamental origin of this relation, and its possible extension beyond semiclassical gravity, remains an open problem.

Third, the fact that shear controls both entropy evolution and geometric instability growth points toward a connection between horizon geometry and information scrambling. Clarifying this relation, in particular in holographic or fully quantum settings, is an important direction.

Finally, it would be valuable to test the present framework in explicit models of quantum gravity, such as Jackiw--Teitelboim gravity, and to explore its implications for global phenomena including the Page curve.

\begin{acknowledgments}
The author gratefully acknowledges the anonymous referee for a careful reading of the manuscript and for several constructive suggestions that helped improve both the presentation and the scope of this work.
\end{acknowledgments}

\appendix

\section{Saturated Dynamics and Attractors}

This appendix is exploratory in nature and analyzes the hypothetical dynamics associated with configurations that exactly saturate the local entropy bound derived above. The entropy Raychaudhuri relation obtained in the main text takes the form of a differential inequality rather than a closed evolution equation. In order to study possible stationary configurations and associated stability properties (see, e.g., \citep{Weinstein2017}), we therefore consider the formal saturating case in which the inequality is replaced by the corresponding equality,
\begin{equation}
\frac{d\Theta}{d\lambda}
=
- \theta \Theta
+
\frac{1}{2}\theta^2
-
\sigma^2.
\label{eq:entropy_ray_eq}
\end{equation}
This construction should be understood only as an effective limiting description of configurations that saturate the bound. No claim is made that generic semiclassical entropy evolution obeys Eq.~\eqref{eq:entropy_ray_eq} exactly. Nevertheless, the saturating case is useful for identifying possible fixed points and for analyzing the local stability structure implied by the bound itself. Stationary (or attractor) configurations are obtained by setting $\frac{d\Theta}{d\lambda} = 0$. Using Eq.~\eqref{eq:entropy_ray_eq}, this gives
\begin{equation}
- \theta \Theta
+
\frac{1}{2}\theta^2
-
\sigma^2 = 0.
\end{equation}
Solving for $\Theta$ yields
\begin{equation}
\Theta_* 
=
\frac{\frac{1}{2}\theta^2 - \sigma^2}{\theta},
\qquad (\theta \neq 0).
\label{eq:Theta_star}
\end{equation}
The sign of $\Theta_*$ depends on the competition between expansion and shear:
\begin{align}
\sigma^2 > \frac{1}{2}\theta^2 
&\;\Rightarrow\; \Theta_* < 0 
\quad \text{(shear-dominated)}, \\
\sigma^2 < \frac{1}{2}\theta^2 
&\;\Rightarrow\; \Theta_* > 0 
\quad \text{(expansion-dominated)}.
\end{align}
Thus, shear shifts the effective attractor toward negative values of the entropy expansion, while expansion drives it toward positive values.

To determine the stability of the fixed point, we consider small perturbations
\begin{equation}
\Theta = \Theta_* + \delta\Theta.
\end{equation}
Inserting this into Eq.~\eqref{eq:entropy_ray_eq} and expanding to linear order in $\delta\Theta$, this yields
\begin{align}
\mathcal{L}_k (\Theta_* + \delta\Theta)
&=
- \theta (\Theta_* + \delta\Theta)
+
\frac{1}{2}\theta^2
-
\sigma^2.
\end{align}
Using the defining equation for $\Theta_*$, the zeroth-order terms cancel, 
leaving
\begin{equation}
\frac{d}{d\lambda}\,\delta\Theta
=
- \theta \,\delta\Theta.
\label{eq:linearized}
\end{equation}
Equation~\eqref{eq:linearized} is a first-order linear differential equation 
with solution
\begin{equation}
\delta\Theta(\lambda)
=
\delta\Theta(0)\, e^{-\theta \lambda}.
\end{equation}
The behavior of perturbations depends on the sign of $\theta$:
\begin{align}
\theta > 0 
&\;\Rightarrow\;
\delta\Theta(\lambda) \to 0
\quad \text{(stable)}, \\
\theta < 0 
&\;\Rightarrow\;
\delta\Theta(\lambda) \to \infty
\quad \text{(unstable)}.
\end{align}
Thus, positive expansion leads to exponential relaxation toward the attractor, while negative expansion leads to instability. In the case of vanishing expansion ($\theta = 0$), Eq.~\eqref{eq:entropy_ray_eq} reduces to
\begin{equation}
\frac{d\Theta}{d\lambda} = -\sigma^2.
\end{equation}
This admits no non-trivial fixed point unless $\sigma = 0$. Instead, the entropy expansion decreases monotonically, corresponding to purely dissipative relaxation without a stationary attractor.

The above analysis shows that the entropy dynamics admits \emph{effective attractor solutions} governed by the interplay between expansion and shear in the saturated case. Shear shifts the attractor toward negative values of $\Theta$, while expansion controls the stability of the fixed point. In this sense, null horizon dynamics exhibits a structure analogous to dissipative dynamical systems, with entropy flow relaxing toward geometry-dependent equilibrium configurations.

\section{Entropy Relaxation Regimes}

We formulate a local entropy evolution based on the entropy Raychaudhuri inequality. Consider a null congruence with tangent vector $k^a$ and generalized expansion  $\Theta$ satisfying
\begin{equation}
\mathcal{L}_k \Theta
\le
- \theta \Theta
+
\frac{1}{2}\theta^2
-
\sigma^2.
\label{eq:EntropyRaychaudhuri_focus}
\end{equation}
We restrict to a shear-dominated (near-stationary) regime in which $|\theta| \ll |\sigma|$, and assume a strictly positive lower bound on the shear, $\sigma^2 \ge \sigma_0^2 > 0$. We further assume that the initial generalized expansion is negative, $\Theta(0) < 0$. Under these conditions, the generalized expansion satisfies
\begin{equation}
\frac{d\Theta}{d\lambda} \le - \sigma_0^2,
\end{equation}
which can be integrated to yield the bound
\begin{equation}
\Theta(\lambda) \le \Theta(0) - \sigma_0^2 \lambda.
\end{equation}
Obviously, the generalized expansion $\Theta$ decreases monotonically along the null generators. In particular, $\Theta(\lambda)$ is driven toward increasingly negative values as $\lambda$ increases, with a rate controlled by the shear. The quantity $\Theta$, which measures the local rate of change of the generalized entropy, is thus damped by the presence of shear. Unlike classical focusing, this evolution does not lead to a finite-time divergence but instead corresponds to a gradual redistribution of entropy along the null congruence. In the absence of shear, $\sigma = 0$, one recovers $\mathcal{L}_k \Theta = 0$, corresponding to a stationary configuration with constant entropy flux.

The Quantum Focussing Conjecture states that $\mathcal{L}_k \Theta \le 0$. The present result provides a dynamical scenario in which this condition is satisfied, identifying $\sigma^2$ as the dominant geometric contribution controlling entropy suppression.

\section{Linear Stability of Entropy Dynamics}

We now analyze small perturbations of the entropy flow around a stationary horizon configuration. We consider a background characterized by
\begin{equation}
\theta_0 = 0, \qquad \sigma_0 = 0, \qquad \Theta_0 = 0,
\end{equation}
and introduce perturbations
\begin{equation}
\Theta = \delta \Theta, \qquad \theta = \delta \theta, \qquad \sigma^2 = \delta \sigma^2.
\end{equation}
To leading order, the entropy Raychaudhuri inequality reduces to
\begin{equation}
\mathcal{L}_k \,\delta \Theta \le - \delta \sigma^2,
\end{equation}
where we have neglected higher-order terms in the perturbations. Since $\delta \sigma^2 \ge 0$, it follows that
\begin{equation}
\mathcal{L}_k \,\delta \Theta \le 0,
\end{equation}
implying that perturbations of the generalized expansion are non-growing. In the presence of anisotropic perturbations, for which $\delta \sigma^2 > 0$, 
one obtains the stricter bound
\begin{equation}
\mathcal{L}_k \,\delta \Theta < 0,
\end{equation}
showing that such perturbations are damped. These results indicate linear stability of the generalized expansion. Shear acts as a dissipative-like mechanism that suppresses anisotropic fluctuations and drives the system toward an isotropic equilibrium.

\section{Physical Timescales}

The evolution equations derived in this work are naturally formulated in terms of the affine parameter $\lambda$ along null generators. However, $\lambda$ does not correspond directly to physical time, since it is invariant under rescalings
\begin{equation}
\lambda \to a \lambda, \qquad k^a \to a^{-1} k^a.
\end{equation}
A physical time coordinate emerges once the null generators are normalized; for stationary horizons, this is typically done using the Killing normalization, which relates the affine parameter to the Killing time $t$ via the surface gravity $\kappa$,
\begin{equation}
\mathcal{L}_k = \kappa \frac{d}{dt}, \qquad \kappa = 2\pi T,
\end{equation}
with $T$ the Hawking temperature. In the near-stationary regime $\theta \simeq 0$, the entropy Raychaudhuri inequality becomes
\begin{equation}
\mathcal{L}_k \Theta \le -\sigma^2,
\end{equation}
and in physical time this is
\begin{equation}
\frac{d\Theta}{dt} \le -\frac{\sigma^2}{\kappa}.
\end{equation}
This suggests a characteristic dissipation-like timescale of the form
\begin{equation}
\tau_{\mathrm{diss}} \sim \frac{\Theta}{|d\Theta/dt|} \sim \frac{\kappa \Theta}{\sigma^2}.
\end{equation}
Assuming, that the generalized expansion $\Theta$ and the shear $\sigma$ are set by comparable geometric scales, $\Theta \sim \sigma$, we find
\begin{equation}
\tau_{\mathrm{diss}} \sim \frac{\kappa}{\sigma}.
\end{equation}
Using the earlier estimate $\lambda_L^{\mathrm{eff}} \sim \sqrt{\sigma^2} \sim \sigma$ for the effective geometric instability exponent, this yields
\begin{equation}
\tau_{\mathrm{diss}} \sim \frac{\kappa}{\lambda_L^{\mathrm{eff}}} \sim \frac{2\pi T}{\lambda_L^{\mathrm{eff}}}.
\end{equation}
Heuristically, this indicates that entropy relaxation occurs on a timescale controlled by the interplay between temperature and the rate of separation of nearby null generators: a stronger separation behavior corresponds to faster entropy relaxation, while lower temperatures suppress the overall dynamics.

Note that the identification $\Theta \sim \sigma$ and $\lambda_L^{\mathrm{eff}} \sim \sigma$ should be regarded as order‑of‑magnitude estimates; the exact relation between the timescale, the geometric instability exponent, and the horizon geometry depends on the details of the background and the state of the quantum fields.

\section{Comparison with Scrambling Scales}

The following comparison is purely heuristic and intended only at the level of scaling relations. It is nevertheless instructive to compare the dissipation timescale with the scrambling time of black holes. Previous work has argued that black holes are fast scramblers, with a characteristic timescale \cite{Sekino2008,Hayden2007}
\begin{equation}
t_* \sim \beta \log S,
\end{equation}
where $\beta = 1/T$ denotes the inverse temperature and $S$ the entropy of the system. We now relate our local dissipation timescale derived above to the global scrambling time. The separation of nearby null generators grows exponentially as
\begin{equation}
X(t) \sim e^{\lambda_L^\mathrm{eff} t},
\end{equation}
indicating that perturbations spread along the horizon with a rate set by the effective geometric instability exponent. Scrambling occurs once the perturbation has spread over all available degrees of freedom, corresponding to
\begin{equation}
e^{\lambda_L^\mathrm{eff} t_{\mathrm{scr}}} \sim S,
\end{equation}
where $S$ is the entropy of the system. This yields
\begin{equation}
t_{\mathrm{scr}} \sim \frac{1}{\lambda_L^\mathrm{eff}} \log S.
\end{equation}
In the case where the geometric instability exponent saturates the maximal bound $\lambda_L = 2\pi T$, this reduces to the standard result from above,
\begin{equation}
t_{\mathrm{scr}} \sim \frac{1}{2\pi T} \log S.
\end{equation}
This derivation suggests that the same geometric quantity controlling the separation of nearby null generators may also be related to global scrambling timescales of the horizon.

\section{Near-Horizon Shear and Surface Gravity}

To connect the geometric instability bound to the horizon temperature, we consider a near-equilibrium, near-horizon expansion of the null generators. For a stationary black hole with surface gravity~$\kappa$, the horizon geometry in Gaussian null coordinates is
\begin{equation}
ds^2 \simeq - 2 \kappa r \, dv^2 + 2 dv dr + \gamma_{AB} dx^A dx^B + \mathcal{O}(r^2),
\end{equation}
where $r$ measures the transverse distance from the horizon. In the near-horizon region, the characteristic tidal scale is controlled by the surface gravity $\kappa$.

Since the shear is sourced by tidal distortions encoded in the Weyl tensor, dimensional analysis suggests that the characteristic shear scale satisfies
\begin{equation}
\sqrt{\sigma^2} = \mathcal{O}(\kappa),
\end{equation}
up to model-dependent corrections and away-from-horizon suppression. The precise relation depends on the specific geometry and perturbations, and should not be regarded as universal. Moreover, in stationary black hole backgrounds, the quantum chaos bound predicts that the Lyapunov exponent is limited by the Hawking temperature $T = \kappa/(2\pi)$,
\begin{equation}
\lambda_L \le 2\pi T,
\end{equation}
as shown by \cite{Maldacena2016}. In our framework, this behavior is qualitatively consistent with the expectation that the characteristic near-horizon shear scale is set by the surface gravity,
\begin{equation}
\sqrt{\sigma^2}
=
\mathcal{O}(\kappa)
\sim
2\pi T,
\end{equation}
in the near-horizon stationary regime, where the dominant geometric instability is controlled by the horizon scale. Combined with the estimate Eq.~\eqref{eq:LyapunovEstimate}, this yields an effective bound of the schematic form
\begin{equation}
\lambda_L^\mathrm{eff} \lesssim 2\pi T,
\end{equation}
in qualitative agreement with the scaling of the standard chaos bound. This suggests that the same geometric structures, controlled by shear and expansion, may underlie both entropy evolution and geometric instability on null horizons.

We emphasize, however, that the relation between $\sigma$, $\kappa$, and the chaos bound is intended only as a scaling argument within the near-horizon stationary regime. A precise identification of $\lambda_L^\mathrm{eff}$ with a microscopic quantum Lyapunov exponent generally requires additional dynamical input from the underlying quantum theory.

\section{Analogy to Hydrodynamics and Dissipation}

The following discussion is heuristic and intended only as an analogy at the level of sign structure and geometric interpretation. The entropy flux inequality
\begin{equation}
\frac{1}{dA} \mathcal{L}_k^2 dS_{\mathrm{gen}}
\le
\frac{1}{4}\left(
\frac{1}{2}\theta^2 - \sigma^2
\right)
\end{equation}
resembles the balance between driving and dissipative contributions familiar from hydrodynamics. In particular, the negative shear contribution $-\sigma^2$ is structurally analogous to dissipative shear terms in viscous fluids, while the positive expansion term $+\frac{1}{2}\theta^2$ acts as a geometric source contribution associated with stretching of the congruence.

This analogy is consistent with the membrane paradigm \citep{Thorne1986,Iqbal2009}, in which black hole horizons behave as effective viscous fluids. Within the present framework, the shear scalar $\sigma^2$ similarly governs the suppression of the generalized entropy flux and therefore plays the role of an effective dissipative-like geometric contribution.

\bibliographystyle{unsrt}
\bibliography{bibliography}

\end{document}